\documentclass[twoside]{article}
\usepackage{fleqn,espcrc2,latexsym,amsmath,amssymb}

\usepackage{graphicx}
\usepackage[figuresright]{rotating}

\renewcommand{\AmS}{{\protect\the\textfont2 A\kern-.1667em\lower.5ex\hbox{M}\kern-.125emS}}

\title{The overlap operator as a continued fraction\thanks{Based on a
talk by U.~Wenger}}

\author{A. Bori\c{c}i\address[MCSD]{Department of Physics and
                          Astronomy, University of Edinburgh,
                          Edinburgh EH9 3JZ, UK},
        A. D. Kennedy\addressmark[MCSD],
        B. J. Pendleton\addressmark[MCSD], and
        U. Wenger\address{Theoretical Physics,
                          Oxford University, 1 Keble Road, Oxford OX1
                          3NP, UK}\thanks{Supported by British PPARC
                          SPG}}
        
\begin{document}

\begin{abstract}
We use a continued fraction expansion of the sign-function in order to
obtain a five dimensional formulation of the overlap lattice Dirac
operator. Within this formulation the inverse of the overlap operator
can be calculated by a single Krylov space method where nested
conjugate gradient procedures are avoided. We show that the five
dimensional linear system can be made well conditioned using
equivalence transformations on the continued fractions. This is of
significant importance when dynamical overlap fermions are simulated. 
\end{abstract}

\maketitle

\section{Introduction}
Let us start with the explicit form of the massive overlap Dirac
operator with mass $\mu$,
\begin{equation}\label{eq:overlap_operator}
D(\mu) = \frac{1}{2} (1 + \mu) + 
  \frac{1}{2} (1 - \mu) \gamma_5 \text{sgn} (H(-m)), 
\end{equation}	
where $H(-m)$ denotes some hermitian Dirac operator with mass $-m<0$
of the order of the inverse lattice spacing.  One method to compute
the action of the sign-function on a vector is by approximating the
function by a rational polynomial \cite{Edwards:1998yw},
\begin{equation}
\text{sgn}(x) \approx \text{sgn}_{\{n,m\}}(x) \equiv 
    \frac{P_{n}(x)}{Q_{m}(x)},
\end{equation}
where $P_n$ and $Q_m$ are irreducible polynomials of degree $n$ and
$m$, respectively. 
The reason for using rational approximations is that for a given
approximation error in the infinity-norm the required degree for
rational approximations is usually much smaller than the one for
polynomial ones, although both converge exponentially in their degree.

There are various methods to perform such an approximation. The
methods differ in how they interpret the notion of the goodness of the
approximation.  One can use any definition of a norm
characterising the deviation of two functions from each other.

The infinity-norm, for example, is defined as the maximal error over a
given interval $[a,b]$. There exists an algorithm (Remez algorithm
\cite{Petrushev}) to find the best (rational) approximation $r(x)$ to a
function $f(x)$ with respect to the infinity-norm (see also the work
of Zolotarjov \cite{Petrushev,Lippert}). That is, the approximation $r(x)$
to a given function $f(x)$ is the rational function of a given degree
that minimises the maximum value of the quantity $|f(x) - r(x)|$ over
the interval $[a,b]$.

In order to approximate the sign-function by a rational polynomial we
choose to find an approximation for a constant odd function over the
interval $[\varepsilon, 1]$. To do so we approximate
$\frac{1}{\sqrt{x}}$ on the interval $[\varepsilon^2, 1]$ to obtain
the rational approximation $r(x)$ and consider the function $x
r(x^2)$. 

The resulting approximation can then be written as 
\begin{equation}
\text{sgn}_{\{2n+1,2m\}}(x) = x c_0
\frac{\prod_{k=1}^{n} (x^2+p_k)}{\prod_{k=1}^{m} (x^2+q_k)}  
\end{equation}  
with real and positive $p_k$ and $q_k$. This factorised form is most
stable for numerical evaluation.

One advantage of using rational polynomials is that one can use the
partial fraction decomposition, e.g.~for $n=m$,
\begin{equation}\label{eq:frac_decomposition}
\text{sgn}_{\{2n+1,2n\}}(x) = x \left(c_0 + 
      \sum_{k=1}^n \frac{c_k}{x^2 + q_k} \right)
\end{equation}
which allows us to compute the action of the sign-function at roughly
 the cost of a single conjugate gradient (CG) inversion of $H^2$
 \cite{Edwards:1998yw,Neuberger:1998my}.

\section{Continued fractions}
It is well known that rational functions are closely related to finite
(truncated) continued fractions which in turn play an important role
in the analytic theory of rational functions
\cite{Wall:1948,Jones:1980}. Indeed, we can write any rational
approximation to the sign-function as a continued fraction
\cite{Neuberger:1999re}
\begin{equation}\label{eq:cont_fraction_expansion}
\text{sgn}_{\{2n+1,2n\}}(x)
	=  \beta_0 x + \cfrac{\alpha_1}{\beta_1 x+ 
			\cfrac{\cdots}{\cdots +
	\cfrac{\alpha_{2n}}{ \beta_{2n} x}}} . 	 
\end{equation}
Then we observe that the inverse of
eq.~(\ref{eq:cont_fraction_expansion}) is just the $(1,1)$-component
of the Schur complement of a larger system
\begin{equation} \label{eq:cont_fraction_matrix}
\left( \begin{array}{ccccc} 
      \beta_0 x & \sqrt{\alpha_1} & &   &    \\ 
  \sqrt{\alpha_1}& -\beta_1 x      &  &   &    \\ 
                 &  & \ddots &  &    \\ 
                 & & & -\beta_{2n-1} x & \sqrt{\alpha_{2n}}\\
	         & & &\sqrt{\alpha_{2n}} &\beta_{2n} x
      \end{array} \right).  
\end{equation}
Instead of inverting (\ref{eq:frac_decomposition}) using a two-level
nested CG(-like) procedure we can now find the inverse of
(\ref{eq:cont_fraction_matrix}), i.e.~its (1,1)-component, in one go,
using just one CG(-like) search in a single higher dimensional
Krylov space \cite{Neuberger:1999re}.

The operation naturally associated with rational functions is the
multiplication and inversion which, of course, is most trivial. For
the overlap operator, eq.~(\ref{eq:overlap_operator}), matters are not
so simple since the additional term proportional to $\gamma_5$ does
not commute with $\text{sgn}(H)$ and its approximations. We can,
however, still make use of some properties of (finite) continued
fractions. The most important one is contained in the following
theorem on equivalence transformations: Two continued fractions
$\{\alpha_0;\{\alpha_1,\beta_1\}, \ldots, \{\alpha_n,\beta_n\}\}$ and
$\{\alpha_0'; \{\alpha_1',\beta_1'\},\ldots, \{\alpha_n',\beta_n'\}\}$
are equivalent if and only if there exists a sequence of non-zero
constants $c_n$ with $c_0 = 1$ such that
\begin{eqnarray} 
\alpha_n'  = & c_n c_{n-1} \alpha_n, & n=1,2,3, \ldots, \nonumber \\
\beta_n'  = & c_n \beta_n, & n=0,1,2, \ldots \, . \nonumber
\end{eqnarray}
This can most easily be seen by writing down the continued fraction
after a transformation, 
\begin{equation}
 \beta_0 + \cfrac{c_1 \alpha_1} {c_1  \beta_1 +
           \cfrac{c_1 c_2 \alpha_2}{c_2 \beta_2+
           \cfrac{\cdots}{\cdots +
           \cfrac{c_{n-1} c_n \alpha_n}{c_n \beta_n}}}} \, . 
\end{equation}
In fact the coefficients $c_i$ parametrise the equivalence class of a
continued fraction corresponding to a given rational polynomial.  It
is important to note that although the equivalence transformation
leaves the value of the continued fraction invariant it affects the
spectrum and correspondingly the condition number of the larger system
(\ref{eq:cont_fraction_matrix}). One might therefore hope to find
within an equivalence class of continued fractions the one with the
smallest condition number.

\section{Application to the overlap operator}
It is easy to apply the equivalence transformation to an overlap
operator in which the sign-function is approximated by a rational
function and expanded as a continued fraction.  In complete analogy to
the previous section we express the linear system of the hermitian
overlap Dirac operator $\gamma_5 D(\mu) \psi = b$ as a simple five
dimensional system \mbox{$H^{(5)}(\mu) \Psi = \chi$} where
\mbox{$\Psi = (\psi,\phi_1,\ldots,\phi_{2n+1})^T$}, \mbox{$\chi =
(b,0,\dots,0)^T$} are five dimensional fermion fields and
$H^{(5)}(\mu)$ is a block tridiagonal matrix similar to
(\ref{eq:cont_fraction_matrix}) with \mbox{$A_+ \gamma_5, - c_1 k_1
H,+c_2 k_2 H, \ldots, - c_{2n+1} k_{2n+1} H$} on the diagonal,
\mbox{$\sqrt{ c_1 A_-}, \sqrt{ c_1 c_2}, \ldots, \sqrt{ c_{2n}
c_{2n+1}}$} with trivial Dirac and colour structure on the off
diagonals and $A_\pm = 1/2(1\pm \mu)$.  The $k_i$'s are uniquely
determined by the rational approximation of the sign-function while
the $c_i$'s are free parameters describing the equivalence class of
continued fractions associated with the given rational approximation.

We observe that an equivalence transformation on the continued
fraction acts in a similar way to block Jacobi preconditioning. Indeed
we can write $H^{(5)}(\mu) = C \cdot H^{(5)}(\mu;c_i=1) \cdot C$ where
$C$ is a block diagonal matrix having the four dimensional unit matrix
${\bf 1}$ in the first block and $\sqrt{c_i} \cdot {\bf 1}, i=1,
\ldots, 2n+1$ in the remaining ones. For the linear system under
consideration this preconditioning comes for free since the first
component of $\Psi$ and $\chi$ are not affected at all.

It turns out that within an equivalence class the condition number of
$H^{(5)}$, $\kappa(H^{(5)})$, can vary by orders of
magnitude. An obvious choice for the equivalence transformation
parameters is $c_i = 1/k_i$, and indeed, with this choice the matrix
is already well-conditioned.  In order to find the optimal values for
the $c_i$ in the sense of giving the lowest $\kappa(H^{(5)})$ one can
study the matrix in the free case where the spectrum can be calculated
analytically. A non-linear minimisation of $\kappa(H^{(5)})$ then
yields a set of optimal values for the coefficients $c_i$.

How much is the effective gain in a practical application?  In order
to address this question we generated an ensemble of pure gauge field
configurations with the Wilson gauge action at $\beta=5.9$ on $6^4$
lattices. We used the standard Wilson Dirac operator $H_W(-m)$ with
$m=1.5$ as the input operator for the overlap and a $n=8$ rational
approximation for the inverse square root on the interval
$[\varepsilon^2, 1]=[0.00724,1]$. This generates a five dimensional
operator with an extent of 18 in the fifth dimension. Then we
calculated the inverse of the hermitian overlap operator $\gamma_5
D(\mu)$ using the five and four dimensional
formulation with different quark masses $\mu$ on a random
vector. 

\begin{figure}[htb]
\vspace{-1.5mm}
    \includegraphics[width=7cm]{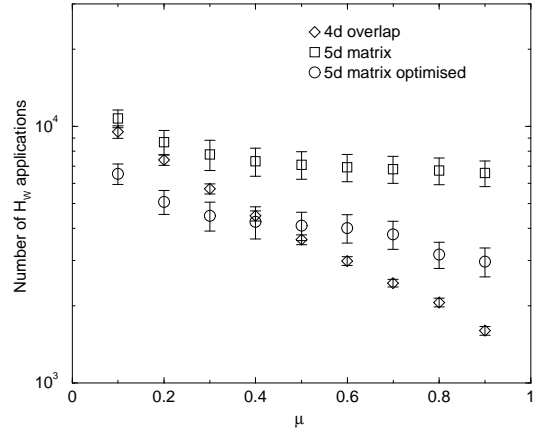}
\vspace{-9mm}
\caption{Number of $H_W$ applications for inverting the overlap
operator in different four and five dimensional representations as a
function of the quark mass $\mu$.}\label{fig:Hw_vs_mu}
\vspace{-4mm}
\end{figure}

Figure \ref{fig:Hw_vs_mu} shows the average number of
$H_W$-matrix multiplications for the inversion in three cases: the
pole approximation, eq.~(\ref{eq:frac_decomposition}), ($4d$ overlap),
the five dimensional operators $H^{(5)}$ with $c_i =1/k_i$ ($5d$
matrix) and $H^{(5)}$ with the optimal $c_i$'s as obtained from the
free field approximation ($5d$ matrix optimised).
The results indicate that it is indeed possible to generate five
dimensional systems which are well conditioned and that the obstacle
of having large condition numbers for the five dimensional system can
be avoided. For small quark masses the inversion seems to behave
better than the one of the corresponding four dimensional system. This
is of particular importance in the context of dynamical overlap
fermion simulations where the inversion of the fermion matrix plays a
crucial role
\cite{Neuberger:1999re,Bode:1999dd,Narayanan:2000qx}. Another
advantage of the five dimensional formulation using continued
fractions is its obvious suitability for parallel computations.


\end{document}